\documentclass[pra,twocolumn,showpacs]{revtex4}
\usepackage{graphics}       % for eps figures
\usepackage{bm}         % for bold math symbols
\usepackage{amsmath}        % for \text and such
\usepackage{latexsym}
\usepackage{amsfonts}
\usepackage{amssymb}

\begin{document}
\title{Retrodiction of  Generalised Measurement Outcomes}
\author{Anthony Chefles}
\affiliation{Department of Physical Sciences, University of
Hertfordshire,
       Hatfield AL10 9AB, Hertfordshire, UK}
\author{Masahide Sasaki}
\affiliation{Communications Research Laboratory, Koganei, Tokyo
184-8795, Japan}

\begin{abstract}
\vspace{0.5in} If a generalised measurement is performed on a
quantum system and we do not know the outcome, are we able to
retrodict it with a second measurement? We obtain a necessary and
sufficient condition for perfect retrodiction of the outcome of a
known generalised measurement, given the final state, for an
arbitrary initial state. From this, we deduce that, when the input
and output Hilbert spaces have equal (finite) dimension, it is
impossible to perfectly retrodict the outcome of any fine-grained
measurement (where each POVM element corresponds to a single Kraus
operator) for all initial states unless the measurement is
unitarily equivalent to a projective measurement. It also enables
us to show that every POVM can be realised in such a way that
perfect outcome retrodiction is possible for an arbitrary initial
state when the number of outcomes does not exceed the output
Hilbert space dimension. We then consider the situation where the
initial state is not arbitrary, though it may be entangled, and
describe the conditions under which unambiguous outcome
retrodiction is possible for a fine-grained generalised
measurement. We find that this is possible for some state if the
Kraus operators are linearly independent. This condition is also
necessary when the Kraus operators are non-singular. From this, we
deduce that every trace-preserving quantum operation is associated
with a generalised measurement whose outcome is unambiguously
retrodictable for some initial state, and also that a set of
unitary operators can be unambiguously discriminated iff they are
linearly independent. We then examine the issue of unambiguous
outcome retrodiction without entanglement.  This has important
connections with the theory of locally linearly dependent and
locally linearly independent operators.
\end{abstract}
 \maketitle

\section{Introduction}
\renewcommand{\theequation}{1.\arabic{equation}}
\setcounter{equation}{0}

One of the most contentious issues in the development of quantum
mechanics was, and continues to be, the measurement process.  The
fact that measurement appears explicitly in the quantum formalism
represents a significant break with the implicit assumption in
classical mechanics that all quantities which enter into the
description of the state of a physical system are observable and
that the measurement process requires no special
treatment\cite{Peres}. It does in quantum mechanics. Among the
consequences of the nature of the quantum measurement process as
expounded by, for example, von Neumann\cite{Von}, are
indeterminism, the impossibility of measuring the state of a
quantum system and uncertainty relations.

However, the projective measurements introduced by von Neumann and
 defined in full generality by L\"{u}ders\cite{Luders} do
retain one significant feature of classical physics. This is the
property of repeatability. Simply stated, if we perform such a
measurement on a  quantum system twice, and if we are able to
reverse any evolution of the state between the measurements, then
the outcome of the second measurement will be the same as that of
the first.

Subsequent developments in quantum measurement theory have shown
that the combination of projective measurements with unitary
interactions leads to a broader range of state transformations and
information-acquisition procedures.  These, which are known as
quantum operations and generalised measurements respectively, are
closely related to each other.

The statistical properties of a generalised measurement are
determined by a set of positive operators forming a positive,
operator-valued measure (POVM).  Generalised measurements enable
one to acquire certain kinds of information about quantum states
which are unobtainable using only projective measurements,
especially if the possible initial states are
non-orthogonal\cite{Holevo73_POM,Helstrom_QDET,Review,Melin}.
However, they do have some disadvantages. One is the fact that
they do not possess the aforementioned repeatability property of
projective measurements. The repeatability of these measurements
is independent of the initial state, which may be arbitrary and
unknown.  It enables us to predict not only the outcome of a
future repetition of the measurement, but also the future
post-measurement state, provided that there is no irreversible
evolution between the measurements. Furthermore, these predictions
will be fulfilled with unit probability.

As well as enabling us to predict the outcome of an identical
measurement, repeatability also enables us to {\em retrodict} the
outcome of a projective measurement and also the post-measurement
state, given that we know which observable was measured and again,
in the absence of subsequent irreversible evolution.

The fact that the repeatability of projective measurements has so
many aspects and consequences suggests that, while these may not
all hold for generalised measurements, some vestiges of
repeatability could be made to hold for these measurements in some
circumstances if we are willing to sacrifice others. This is the
issue we investigate in this paper. The particular aspect of the
repeatability of projective measurements we would like to retain
is outcome retrodictability. As one might expect, when this is
possible, the measurement which carries out the retrodiction will,
in general, differ from the original measurement in this wider
context.

%---------------------------------
%

%
%-------<Added : Sasaki>-------
%
%     the reader would need further explanation of our context
%     as I was confused at the first reading of this draft.
%     Instead we may shorten the orher paragraphs.
%
It is well-established that the implementation of a generalised
measurement will often involve a projective measurement on an
extended space \cite{Helstrom_QDET,Peres}, for example, a
projective measurement on a Cartesian product (Naimark) extension
or a unitary-projection scheme on a direct, or tensor product
extension. However, it is typically the case that we do not have
access to this extension, which is assumed to be the case
throughout this paper.  When we address the issue of measurement
outcome retrodictability, the retrodiction operators will act only
on the space of system post-measurement states and not on such an
extension. We shall, however, allow for the possibility that the
space of post-measurement states differs from that of the
preparation states whenever making this distinction is necessary
for a fully general analysis.

We should also emphasise the  distinction between the idea of
retrodicting the outcome of a generalised measurement and the
formalism of retrodictive quantum mechanics.  The latter was
proposed originally by Aharonov and coworkers\cite{Aharonov} and
has recently been extended and applied in numerous interesting
ways by Barnett and coworkers\cite{Steve}. In retrodictive quantum
mechanics, the aim is to use accessible measurement data to
retrodict the initial state of a quantum system.  The retrodicted
information is then quantum information. In the present context,
although a measurement has been carried out, the result is not
accessible and it is this classical measurement result that we aim
to retrodict.

Our motivation for focusing on this particular aspect of
repeatability is as follows: if we know the result of a known
measurement then in practical situations we would seldom have any
reason to carry it out again.  The issue of repeatability, or
non-repeatability, will be important in situations when a
measurement has been performed and the result has been lost or
otherwise made inaccessible to us. If we do not know the
measurement result then, in the most favourable scenario, we will
have access to the final state. This is a mixture of the
post-measurement states corresponding to the various possible
outcomes weighted by their respective probabilities. When we do
have access to the system following the measurement, which we
shall assume to be the case, we will be concerned with how its
state has been transformed by the measurement process. If the
initial state is represented by a density operator ${\rho}$, then
the final state will be obtained by a completely positive, linear,
trace-preserving (CPLTP) map
${\Phi}:{\rho}{\rightarrow}{\Phi}({\rho})$.

In projective measurements, repeatability and thus perfect outcome
retrodiction are possible for an arbitrary, unknown, initial
state. At the outset, we make a distinction between two kinds of
generalised measurement: fine-grained and coarse-grained
measurements.  These correspond, respectively, to situations where
each POVM element is related to a single, or multiple Kraus
transformation operators. The former is clearly a special case of
the latter.  Section II is devoted to the examination of perfect
outcome retrodiction, that is, deterministic, error-free
retrodiction of the outcome of a known generalised measurement. We
derive a necessary and sufficient condition for such perfect
retrodiction to be possible for an arbitrary initial state and
show that there is no advantage to be gained if the initial state,
though arbitrary, is known.   The remainder of this section is
devoted to unravelling some implications of this condition.  We
show that it implies that, when the input and output Hilbert
spaces have equal dimension, the only fine-grained measurements
with perfectly retrodictable outcomes for arbitrary initial states
are those which are unitarily equivalent to projective
measurements. However, we also show that there exists a large
class of coarse-grained generalised measurements which are highly
dissimilar to projective measurements for which perfect outcome
retrodiction, with an arbitrary initial state, is possible.   We
show that a necessary and sufficient condition for a particular
POVM to have an associated, typically coarse-grained, generalised
measurement whose outcome is perfectly retrodictable for all
initial states is that the number of outcomes does not exceed the
dimension of the output Hilbert space of the system. We also show
how such measurements can be realised in terms of the
unitary-projection picture of generalised measurements.

In section III we drop the condition of perfect retrodiction and
require instead that the outcome can be retrodicted,
unambiguously, with some probability instead. We also, for the
most part, drop the condition that the initial state may be
arbitrary, and require only that the outcome is retrodictable for
at least one known, initial state.  We focus on fine-grained
measurements and allow for the possibility of the system being
initially entangled with an additional, ancillary system. We show
that, when such entanglement is permitted, the measurement
operations for which this is possible are closely related to the
`canonical' representations of general quantum operations, first
studied by Choi\cite{Choi}. These canonical representations have
linearly independent Kraus operators.   We find that a general
sufficient and, for `finite-strength' measurements\cite{Fuchs},
which, in the fine-grained case, have non-singular Kraus
operators, necessary condition for unambiguous retrodiction of the
outcome of a fine-grained generalised measurement for some,
possibly entangled, initial state is that the Kraus operators are
linearly independent.  Every CPLTP map has a Choi canonical
representation, and so every trace-preserving quantum operation
has an associated, fine-grained, generalised measurement amenable
to unambiguous outcome retrodiction. A further consequence of our
analysis, relating to unitary operator discrimination, is that a
necessary and sufficient condition for unambiguous discrimination
among a set of unitary operators is that they are linearly
independent.

We finally examine the issue of unambiguous outcome retrodiction
without entanglement.  We focus on finite-strength, fine-grained
measurements. For such measurements, we find that a necessary and
sufficient condition for unambiguous outcome retrodiction for some
non-entangled initial pure state is that the Kraus operators are
not {\em locally} linearly dependent.  We use this, together with
some results recently obtained by \v{S}emrl and
coworkers\cite{Semrl1,Semrl2} relating to locally linearly
dependent operators, to explore the relationship between
unambiguous outcome retrodiction without entanglement and local
linear dependence.  We then explore the possibility of unambiguous
outcome retrodiction for every initial, pure, separable state. For
fine-grained, finite-strength measurements, we find that this is
possible only when the Kraus operators are locally linearly
independent.

\section{Perfect outcome retrodiction for arbitrary initial states}
\renewcommand{\theequation}{2.\arabic{equation}}
\setcounter{equation}{0}

Consider a quantum system ${\cal Q}$.  Its initial state lies in a
Hilbert space which we will denote by ${\cal H}_{\cal Q}$.  Except
where explicitly stated otherwise, this will have finite dimension
$D_{\cal Q}$. A generalised measurement ${\cal M}_{\cal Q}$ is
carried out on this system.  We assume that the number of possible
outcomes of this measurement is also finite and shall denote this
by $N$.

 The possible outcomes of the measurement ${\cal M}_{\cal Q}$ will be
labelled by the index $k{\in}\{1,{\ldots},N\}$. Associated with
the $k$th outcome is a linear, positive, quantum detection
operator, or positive operator-valued measure (POVM) element
${\Pi}_{k}:{\cal H}_{\cal Q}{\rightarrow}{\cal H}_{\cal Q}$. These
satisfy
\begin{equation}
\sum_{k=1}^{N}{\Pi}_{k}=1_{\cal Q},
\end{equation}
where $1_{\cal Q}$ is the identity operator on ${\cal H}_{\cal
Q}$. The probability of outcome $k$ when the initial state is
described by the density operator ${\rho}$ is
\begin{equation}
P(k|{\rho})=\mathrm{Tr}({\Pi}_{k}{\rho}).
\end{equation}
Suppose that the measurement ${\cal M}_{\cal Q}$ is carried out on
${\cal Q}$ and that the outcome is withheld from us.  We do,
nevertheless, have access to the final state of the system.  On
the basis of this, can we retrodict the measurement outcome?

To proceed, we must account for the manner in which the state of
the system is transformed by the measurement process.  Let
$\tilde{\cal H}_{\cal Q}$ be the Hilbert space of post-measurement
states.    These definitions enable us to allow for the
possibility that the initially prepared system and the system
corresponding to the space of post-measurement states, which will
subsequently be subjected to a retrodiction attempt, may be
different.  For example, the initial state may be that of an atom,
yet the final state that of an electromagnetic field mode.
However, for the sake of notational convenience, we shall denote
both the initially prepared system and the final, interrogated
system by the symbol ${\cal Q}$, as it will be clear from the
context which system is being referred to.

We distinguish between two kinds of generalised measurement. We
will refer to these as fine-grained measurements and
coarse-grained measurements. In a fine-grained measurement,
corresponding to each detection operator ${\Pi}_{k}$, there is a
single Kraus operator $A_{k}:{\cal H}_{\cal
Q}{\rightarrow}\tilde{\cal H}_{\cal Q}$ such that
\begin{equation}
{\Pi}_{k}=A^{\dagger}_{k}A_{k}
\end{equation}
and the final, normalised state of the system when the outcome is
$k$ is given by the transformation
\begin{equation}
{\rho}{\rightarrow}{\rho}_{k}=\frac{A_{k}{\rho}A^{\dagger}_{k}}{P(k|{\rho})}.
\end{equation}

In a coarse-grained measurement, corresponding to the operator
${\Pi}_{k}$, there is a set of $R$ Kraus operators $A_{kr}$, where
$r{\in}\{1,{\ldots},R\}$, some of which may be zero, such that
\begin{equation}
{\Pi}_{k}=\sum_{r=1}^{R}A^{\dagger}_{kr}A_{kr}.
\end{equation}
The final, normalised state of the system when the outcome is $k$
is given by the transformation
\begin{equation}
{\rho}{\rightarrow}{\rho}_{k}=\frac{\sum_{r=1}^{R}A_{kr}{\rho}A^{\dagger}_{kr}}{P(k|{\rho})}
\end{equation}
where, in both cases, $P(k|{\rho})$ is given by Eq. (2.2).  We can
easily see from these definitions that fine-grained measurements
are a special case of coarse-grained measurements.

Given the post-measurement system, to retrodict the measurement
outcome we must be able to distinguish between the $k$ possible
post-measurement states ${\rho}_{k}$. We will say that the
retrodiction is perfect if the probability of error is zero and
the retrodiction is deterministic, i.e. the probability of the
attempt at retrodiction giving an inconclusive result is also
zero. Perfect retrodiction will be possible only if the
${\rho}_{k}$ are orthogonal, that is
\begin{equation}
{\mathrm{Tr}}({\rho}_{k'}{\rho}_{k})=\mathrm{Tr}({\rho}_{k}^{2}){\delta}_{kk'},
\end{equation}
or equivalently, that
\begin{equation}
{\rho}_{k'}{\rho}_{k}={\rho}_{k}^{2}{\delta}_{kk'}.
\end{equation}
Even if, for every initial state ${\rho}$, the final states
${\rho}_{k}$ are orthogonal, it could be the case that a different
measurement is required to distinguish between the final states
for each initial state.  So, it would appear that there are two
distinct cases to consider when examining the issue of whether the
outcome of a generalised measurement can be perfectly retrodicted
for an arbitrary initial state, corresponding to whether the
initial state is known or unknown.  The former case is clearly at
least as favourable as the latter, since in the former there is
the possibility of tailoring the retrodicting measurement to suit
the possible final states, and by implication the initial state,
which we cannot do in the latter case. It follows that if perfect
retrodiction of the outcome of a generalised measurement ${\cal
M}_{\cal Q}$ is possible for an arbitrary, known, initial state,
then it is also possible if the initial state is unknown. The
following theorem gives a necessary and sufficient condition for
perfect outcome retrodiction for all initial states, and moreover
shows that there is, in fact, no advantage to be gained when the
initial state, though arbitrary, is known.

\newtheorem{theorem}{Theorem}
\begin{theorem}
A quantum system ${\cal Q}$ is initially prepared in the state
$|{\psi}{\rangle}{\in}{\cal H}_{\cal Q}$.  A generalised
measurement ${\cal M}_{\cal Q}$ with $N$ POVM elements ${\Pi}_{k}$
 and Kraus operators $A_{kr}$ satisfying Eq. (2.5) is carried out
on ${\cal Q}$.  The Hilbert space of the post-measurement states
$\tilde{\cal H}_{\cal Q}$ has dimension $\tilde{D}_{\cal Q}$.  A
necessary and sufficient condition for the outcome of ${\cal
M}_{\cal Q}$ to be perfectly retrodictable for every initial state
$|{\psi}{\rangle}{\in}{\cal H}_{\cal Q}$ is
\begin{equation}
A_{k'r'}^{\dagger}A_{kr}={\delta}_{kk'}A_{kr'}^{\dagger}A_{kr},
\end{equation}
for all $r,r'{\in}\{1,{\ldots},R\}$ and irrespective of whether or
not $|{\psi}{\rangle}$ is known.
\end{theorem}
\noindent $\mathbf{Proof}$.  We will prove this theorem by
establishing the necessity of condition (2.9) when the initial
state is arbitrary and known.  Subsequently, we will show that
this condition is sufficient when the initial state is arbitrary
and unknown. Thus, knowing the state confers no benefits in the
context of this problem.   To prove necessity, we will make use of
the unnormalised final density operators
\begin{equation}
{\tilde
\rho}_{k}=\sum_{r=1}^{R}A_{kr}|{\psi}{\rangle}{\langle}{\psi}|A^{\dagger}_{kr}.
\end{equation}
We do this to avoid unnecessary complications which arise when the
probability of one of the outcomes is zero.  When this is so, the
corresponding unnormalised final density operator will also be
zero, but shall see that this causes no problems.

From Eq. (2.7), we see that the necessary condition for perfect
outcome retrodiction given the initial state $|{\psi}{\rangle}$ is
\begin{equation}
{\mathrm{Tr}}({\tilde \rho}_{k'}{\tilde \rho}_{k})=0,
\end{equation}
when $k{\neq}k'$ and for all $|{\psi}{\rangle}{\in}{\cal H}_{\cal
Q}$. This is the sole condition for perfect retrodictability we
will impose in order to establish the necessity of Eq. (2.9). It
says that the final states are orthogonal, which must be true if
we can distinguish between them perfectly (using a projective
measurement.) We do not require that the same distinguishing
measurement is suitable for all initial states, so we take the
initial state to be known, and assume that the appropriate
distinguishing measurement can always be carried out.

Substituting (2.10) into (2.11), we find that
\begin{eqnarray}
&\mathrm{Tr}&\left(\sum_{r,r'=1}^{R}A_{k'r'}|{\psi}{\rangle}{\langle}{\psi}|A^{\dagger}_{k'r'}A_{kr}|{\psi}{\rangle}{\langle}{\psi}|A^{\dagger}_{kr}\right),
\nonumber
\\&=&\sum_{r,r'=1}^{R}|{\langle}{\psi}|A^{\dagger}_{k'r'}A_{kr}|{\psi}{\rangle}|^{2}=0,
\end{eqnarray}
for $k{\neq}k'$.  From this, we see that
\begin{eqnarray}
&&{\langle}{\psi}|A^{\dagger}_{k'r'}A_{kr}|{\psi}{\rangle}={\delta}_{kk'}{\langle}{\psi}|A^{\dagger}_{kr'}A_{kr}|{\psi}{\rangle}
,\nonumber \\
&{\Rightarrow}&{\langle}{\psi}|\left(A^{\dagger}_{k'r'}A_{kr}-{\delta}_{kk'}A^{\dagger}_{kr'}A_{kr}\right)|{\psi}{\rangle}=0
\end{eqnarray}
for all $r,r'{\in}\{1,{\ldots},R\}$ and all
$|{\psi}{\rangle}{\in}{\cal H}_{\cal Q}$, which implies Eq. (2.9).
This proves necessity.

We now prove that Eq. (2.9) is a sufficient condition for perfect
outcome retrodiction when the initial state is both arbitrary and
unknown. We show that there exists a projective measurement which
is independent of the initial state and can be used to distinguish
perfectly between the post-measurement states ${\rho}_{k}$.
Consider the following subspaces of $\tilde{\cal H}_{\cal Q}$:
\begin{equation}
\tilde{\cal H}_{{\cal
Q}k}=\mathrm{span}\left\{|{\phi}{\rangle}{\in}\tilde{\cal H}_{\cal
Q}:{\langle}{\phi}|\left(\sum_{r=1}^{R}A_{kr}A^{\dagger}_{kr}\right)|{\phi}{\rangle}>0,
\right\}
\end{equation}
that is, $\tilde{\cal H}_{{\cal Q}k}$ is the support of the
operator $\sum_{r}A_{kr}A^{\dagger}_{kr}:\tilde{\cal H}_{{\cal
Q}}{\rightarrow}\tilde{\cal H}_{{\cal Q}}$.  Let
$P_{k}:\tilde{\cal H}_{{\cal Q}}{\rightarrow}\tilde{\cal H}_{{\cal
Q}}$ be the projector onto $\tilde{\cal H}_{{\cal Q}k}$.  We will
prove that when Eq. (2.9) is satisfied, these projectors are
orthogonal and form a projective measurement which can always be
used to distinguish perfectly between the post-${\cal M}_{\cal Q}$
states.

To show that they form a projective measurement, define
\begin{equation}
G_{k}=\sum_{r=1}^{R}A_{kr}A^{\dagger}_{kr}.
\end{equation}
Eq. (2.9) implies that
\begin{equation}
G_{k}G_{k'}={\delta}_{kk'}G_{k}^{2}.
\end{equation}
It follows from this, and the positivity of the $G_{k}$, that,
when $k{\neq}k'$, every eigenvector of $G_{k}$ corresponding to a
non-zero eigenvalue is orthogonal to every eigenvector of $G_{k'}$
corresponding to a non-zero eigenvalue. Let ${\tilde {\cal
H}}_{G}$ be the support of the operator
$\sum_{kr}A_{kr}A^{\dagger}_{kr}$, having dimension $D_{G}$. It
follows from Eq. (2.16) that $\tilde{\cal H}_{G}$ has an
orthonormal basis $\{|g_{j}{\rangle}\}$ in terms of which we can
write
\begin{equation}
G_{k}=\sum_{j=1}^{D_{G}}g_{jk}|g_{j}{\rangle}{\langle}g_{j}|,
\end{equation}
where
\begin{equation}
g_{jk}g_{j'k'}={\delta}_{kk'}g_{jk}g_{j'k}\;{\forall}j,j'.
\end{equation}
It follows from Eq. (2.17) that
\begin{equation}
P_{k}=\sum_{j:g_{jk}{\neq}0}|g_{j}{\rangle}{\langle}g_{j}|.
\end{equation}
Making use of Eq. (2.18), we see that these projectors are
orthogonal, i.e
\begin{equation}
P_{k}P_{k'}={\delta}_{kk'}P_{k}.
\end{equation}
They are also complete on the space $\tilde{\cal H}_{G}$. To prove
that a projective measurement based on these projectors can
distinguish perfectly between the ${\rho}_{k}$, we make use of the
fact that the support of ${\tilde\rho}_{k}$ is a subspace of
$\tilde{\cal H}_{{\cal Q}k}$. To prove this, we make use of the
fact that the positivity of $1_{\cal Q}-{\rho}$ implies that
\begin{equation}
\tilde{\rho}_{k}{\leq}G_{k}.
\end{equation}
In other words,
\begin{equation}
{\langle}{\phi}|\tilde{\rho}_{k}|{\phi}{\rangle}{\leq}{\langle}{\phi}|G_{k}|{\phi}{\rangle}\;{\forall}|{\phi}{\rangle}{\in}\tilde{\cal
H}_{\cal Q}.
\end{equation}
Hence, every state $|{\phi}{\rangle}$ which is in the support of
$\tilde{\rho}_{k}$ is also in $\tilde{\cal H}_{{\cal Q}k}$, the
support of $G_{k}$.  Furthermore, for any final state ${\rho}_{k}$
with non-zero outcome probability, the support of ${\rho}_{k}$ is
the same as that of $\tilde{\rho}_{k}$. The fact that the
subspaces $\tilde{\cal H}_{{\cal Q}k}$ are orthogonal and can thus
be perfectly distinguished using a projective measurement on the
space ${\tilde {\cal H}}_{{\cal Q}}$ based on the projectors
$P_{k}$ enables us to distinguish between the states ${\rho}_{k}$
with the same projective measurement, irrespective of the initial
state
$|{\psi}{\rangle}$. This completes the proof${\Box}$.\\

The fact that (2.19) is a sufficient condition for perfect outcome
retrodiction when $|{\psi}{\rangle}$ is an arbitrary, unknown pure
state $|{\psi}{\rangle}$ can easily be seen to imply that it is
also sufficient when the initial state is an arbitrary mixed state
${\rho}$.

Theorem 1 implies the following for fine-grained measurements:

\begin{theorem}
A quantum system ${\cal Q}$ is initially prepared in the state
$|{\psi}{\rangle}{\in}{\cal H}_{\cal Q}$.  A fine-grained
generalised measurement ${\cal M}_{\cal Q}$ is carried out on
${\cal Q}$. If $\tilde{D}_{\cal Q}=D_{\cal Q}$, the outcome of
${\cal M}_{\cal Q}$  is perfectly retrodictable for all
$|{\psi}{\rangle}{\in}{\cal H}_{\cal Q}$, irrespective of whether
or not $|{\psi}{\rangle}$ is known, if and only if ${\cal M}_{\cal
Q}$ is a projective measurement followed by a unitary
transformation from ${\cal H}_{\cal Q}$ to $\tilde{\cal H}_{\cal
Q}$, that is
\begin{equation}
{\Pi}_{k'}{\Pi}_{k}={\delta}_{kk'}{\Pi}_{k},
\end{equation}
where each POVM element is related to its corresponding Kraus
operator in the following way:
\begin{equation}
A_{k}=U{\Pi}_{k}.
\end{equation}
and $U$ is a unitary transformation from ${\cal H}_{\cal Q}$ to
$\tilde{\cal H}_{\cal Q}$.
\end{theorem}
\noindent $\mathbf{Proof}$. For a fine-grained measurement, we see
that it follows from Eq. (2.9) that a necessary and sufficient
condition for perfect outcome retrodiction with an arbitrary,
known or unknown, initial state $|{\psi}{\rangle}{\in}{\cal
H}_{\cal Q}$ is
\begin{equation}
A_{k'}^{\dagger}A_{k}={\delta}_{kk'}A_{k}^{\dagger}A_{k}.
\end{equation}
Sufficiency is easily proven.  When Eqs. (2.23) and (2.24) are
satisfied, we see that
$A_{k'}^{\dagger}A_{k}={\Pi}_{k'}{\Pi}_{k}={\delta}_{kk'}{\Pi}_{k}={\delta}_{kk'}A_{k}^{\dagger}A_{k}$.
This proves sufficiency.  To prove necessity, we notice that, for
fine-grained measurements, Eqs. (2.3) and (2.9) imply
\begin{equation} A^{\dagger}_{k'}A_{k}={\Pi}_{k}{\delta}_{kk'}.
\end{equation}
If we sum both sides of this with respect to $k$ and $k'$, and
make use of the resolution of the identity (2.1), we find that
\begin{equation}
\left(\sum_{k'=1}^{N}A^{\dagger}_{k'}\right)\left(\sum_{k=1}^{N}A_{k}\right)=1_{\cal
Q},
\end{equation}
which implies that $\sum_{k=1}^{N}A_{k}$ is an isometry, which, if
$\tilde{D}_{\cal Q}=D_{\cal Q}$, is necessarily unitary. We will
write
\begin{equation}
\sum_{k=1}^{N}A_{k}=U.
\end{equation}
Summing both sides of Eq. (2.26) over $k'$, and making use of the
adjoint of Eq. (2.28), we obtain
\begin{equation}
A_{k}=U{\Pi}_{k}.
\end{equation}
Substituting this into Eq. (2.26) gives
\begin{equation}
{\Pi}_{k'}{\Pi}_{k}={\Pi}_{k}{\delta}_{kk'}.
\end{equation}
So, the POVM elements of the measurement ${\cal M}_{\cal Q}$ form
a set of orthogonal projectors.  Thus, if perfect retrodiction of
the outcome of a fine-grained measurement is possible for every
initial state, even if the actual state is known, then when the
input and output Hilbert spaces have the same dimension, the
measurement is a projective measurement followed by a unitary
transformation. This completes the
proof${\Box}$.\\

It is natural to examine in more detail the issue of outcome
retrodictability for more general, coarse-grained measurements. As
we shall see, there do exist coarse-grained measurements which are
highly dissimilar to projective measurements for which perfect
outcome retrodiction is possible. Prior to showing this, we make
the following observation which will put our findings in context.
The statistical properties of a generalised measurement are
determined solely by the POVM elements ${\Pi}_{k}$.  These
operators can always be decomposed in the manner of (2.5).  This
decomposition is non-unique, so a POVM with elements ${\Pi}_{k}$
defines an {\em equivalence class} ${\cal E}(\{{\Pi}_{k}\})$ of
measurements, each element of which corresponds to a particular
coarse-grained operator-sum decomposition of the form (2.6) with
fine-grained decompositions being special cases. Having these
ideas in mind, we can ask the following question: under what
circumstances does the equivalence class associated with a
particular POVM contain a generalised measurement whose outcome is
perfectly retrodictable for an arbitrary pure initial state? For
generalised measurements with a finite number of outcomes, this is
answered by the following theorem:

\begin{theorem}
Let ${\cal E}(\{{\Pi}_{k}\})$ be the equivalence class of
generalised measurements associated with a particular POVM with
$N<{\infty}$ elements ${\Pi}_{k}$, where these operators act on
the Hilbert space ${\cal H}_{\cal Q}$ of a quantum system ${\cal
Q}$. This space has dimension ${\cal D}_{\cal Q}$.  The Hilbert
space of the post-measurement states, $\tilde{\cal H}_{\cal Q}$,
has dimension $\tilde{D}_{\cal Q}$.  A necessary and sufficient
condition for the existence of a measurement ${\cal M}_{\cal
Q}{\in}{\cal E}(\{{\Pi}_{k}\})$ whose outcome is perfectly
retrodictable for an arbitrary pure initial state is

\begin{equation}
N{\leq}\tilde{D}_{\cal Q}.
\end{equation}
\end{theorem}
\noindent $\mathbf{Proof}$.  To prove necessity we make use of the
fact that for every generalised measurement with $N<{\infty}$
outcomes, there exists a state vector $|{\psi}{\rangle}{\in}{\cal
H}_{\cal Q}$ such that
$P(k|{\psi})>0\;\;{\forall}\;\;k{\in}\{1,{\ldots},N\}$. To see why
this is so, let ${\cal K}_{k}$ be the kernel of ${\Pi}_{k}$. None
of the ${\Pi}_{k}$ are equal to the zero operator, so the space
${\cal K}_{k}$ is at most $D_{\cal Q}-1$ dimensional. It follows
that if there is no vector $|{\psi}{\rangle}{\in}{\cal H}_{\cal
Q}$ such that
${\langle}{\psi}|{\Pi}_{k}|{\psi}{\rangle}>0\;\;{\forall}\;\;k{\in}\{1,{\ldots},N\}$,
then every $|{\psi}{\rangle}{\in}{\cal H}_{\cal Q}$ is an element
of at least one of the ${\cal K}_{k}$.  We conclude that ${\cal
H}_{\cal Q}={\cup}_{k=1}^{N}{\cal K}_{k}$. This statement, that
the $D_{\cal Q}$ dimensional Hilbert space ${\cal H}_{\cal Q}$ is
the union of a finite set of Hilbert spaces of strictly lower
dimension, is clearly false.  For example, a two dimensional plane
is not the union of a finite set of one dimensional rays. Hence,
for each generalised measurement with a finite number of potential
outcomes, there exists a pure initial state for which all of these
outcomes have non-zero probability of occurrence \cite{footnote1}.

Suppose that ${\cal Q}$ is initially prepared such a state. The
final state corresponding to the $k$th outcome is ${\rho}_{k}$. If
Eq. (2.31) is not satisfied, then the number of final states will
exceed the dimension $\tilde{D}_{\cal Q}$ of $\tilde{\cal H}_{\cal
Q}$. To retrodict the outcome of the measurement perfectly, we
must be able to distinguish between the states ${\rho}_{k}$
perfectly. The supports of these states must be orthogonal, which
is clearly impossible if their number exceeds the dimension of
$\tilde{\cal H}_{\cal Q}$. This proves necessity.

We will prove sufficiency constructively, which is to say that we
will explicitly derive a measurement in the equivalence class
corresponding to any POVM which satisfies Eq. (2.31) for which the
outcome is perfectly retrodictable for an arbitrary pure initial
state. To begin, we write the ${\Pi}_{k}$ in spectral
decomposition form
\begin{equation} {\Pi}_{k}=\sum_{r=1}^{D_{\cal Q}}{\pi}_{kr}|{\pi}_{kr}{\rangle}{\langle}{\pi}_{kr}|,
\end{equation}
where the ${\pi}_{kr}$ are real and non-negative and, for each
$k$, the set $\{|{\pi}_{kr}{\rangle}\}$ is an orthonormal basis
for ${\cal H}_{\cal Q}$. We require a set of Kraus operators
$A_{kr}:{\cal H}_{\cal Q}{\rightarrow}\tilde{\cal H}_{\cal Q}$
satisfying
\begin{equation}
{\Pi}_{k}=\sum_{r=1}^{D_{\cal Q}}A^{\dagger}_{kr}A_{kr},
\end{equation}
for the ${\Pi}_{k}$ defined by Eq. (2.32) and which satisfy the
perfect retrodiction condition in Eq. (2.9).  To this end,
consider
\begin{equation}
A_{kr}=\sqrt{{\pi}_{kr}}|x_{k}{\rangle}{\langle}{\pi}_{kr}|,
\end{equation}
where the set $\{|x_{k}{\rangle}\}$ is any set of $N$ orthonormal
states in $\tilde{\cal H}_{\cal Q}$. Notice that this construction
is possible only if (2.31) is satisfied. The orthonormality of the
$|x_{k}{\rangle}$ implies that the $A_{kr}$ satisfy the perfect
outcome retrodictability condition Eq. (2.9). One can also easily
verify that they are
related to the ${\Pi}_{k}$ in Eq. (2.33) through Eq. (2.34).  This completes the proof${\Box}$.\\

The forgoing discussion has been somewhat abstract.  It would be
helpful to have a concrete physical understanding of how these
measurements can be implemented.  Generalised measurements are
commonly understood as resulting from a unitary interaction with
an ancillary system, followed by a projective measurement on the
latter.  For $\tilde{D}_{\cal Q}=D_{\cal Q}$, we shall see here
how to form a unitary-projection implementation of any POVM which
satisfies Eq. (2.31) whose outcome is perfectly retrodictable
given what we shall shortly refer to as a standard implementation.

We begin with the following well-known fact about generalised
measurements, as described, for example, by Kraus\cite{Kraus}.
Suppose that we have a POVM ${\Pi}_{k}$, with
$k{\in}\{1,{\ldots},{\cal D}_{\cal Q}\}$ which we wish to measure.
This POVM may be factorised as
\begin{equation}
{\Pi}_{k}=B^{\dagger}_{k}B_{k}.
\end{equation}
for some operators $B_{k}:{\cal H}_{\cal
Q}{\rightarrow}\tilde{\cal H}_{\cal Q}$.  Let us introduce a
${\cal D}_{\cal Q}$ dimensional ancilla ${\cal A}_{1}$ with
Hilbert space ${\cal H}_{{\cal A}_{1}}$, initially prepared in the
state $|{\chi}{\rangle}$. For any operators $B_{k}$ satisfying Eq.
(2.35) and the resolution of the identity (2.1), there exists a
unitary transformation $U_{{\cal Q}{\cal A}}:{\cal H}_{\cal
Q}{\otimes}{\cal H}_{{\cal A}_{1}}{\rightarrow}\tilde{\cal
H}_{\cal Q}{\otimes}{\cal H}_{{\cal A}_{1}}$ such that
\begin{equation}
U_{{\cal Q}{\cal A}_{1}}|{\psi}{\rangle}_{\cal
Q}{\otimes}|{\chi}{\rangle}_{{\cal A}_{1}}=\sum_{k=1}^{{\cal
D}_{\cal Q}}(B_{k}|{\psi}{\rangle})_{\cal
Q}{\otimes}|x_{k}{\rangle}_{{\cal A}_{1}},
\end{equation}
where the $\{|x_{k}{\rangle}\}$ is an orthonormal basis set for
${\cal H}_{{\cal A}_{1}}$.  A measurement on ${\cal A}_{1}$ in
this basis, yielding the result $k$, transforms the state of
${\cal Q}$ from $|{\psi}{\rangle}$ into
$B_{k}|{\psi}{\rangle}/\sqrt{P(k|{\psi})}$, with probability
$P(k|{\psi})={\langle}{\psi}|{\Pi}_{k}|{\psi}{\rangle}$.  We will
refer to this construction as a standard implementation of a POVM.

To obtain from this measurement a perfectly retrodictable one
which is also in the equivalence class of the same POVM, we
introduce a further ancilla ${\cal A}_{2}$ with $\tilde{\cal
D}_{\cal Q}$ dimensional Hilbert space ${\cal H}_{{\cal A}_{2}}$,
also initially prepared in the state $|{\chi}{\rangle}$. Following
the action of $U_{{\cal Q}{\cal A}}$, we apply a unitary copying
transformation on ${\cal A}_{1}{\cal A}_{2}$ which perfectly
copies the orthogonal states $|x_{k}{\rangle}$, that is,
\begin{equation}
\mathrm{COPY}_{{\cal A}_{1}{\cal A}_{2}}|x_{k}{\rangle}_{{\cal
A}_{1}}{\otimes}|{\chi}{\rangle}_{{\cal
A}_{2}}=|x_{k}{\rangle}_{{\cal
A}_{1}}{\otimes}|x_{k}{\rangle}_{{\cal A}_{2}}.
\end{equation}
Since $\tilde{D}_{\cal Q}=D_{\cal Q}$, we can carry out the SWAP
operation on ${\cal Q}{\cal A}_{1}$, which exchanges the states of
these two systems.  The entire unitary interaction between ${\cal
Q}$ and the ancilla ${\cal A}_{1}{\cal A}_{2}$ is then
\begin{eqnarray}
&&\mathrm{SWAP}_{{\cal Q}{\cal A}_{1}}\mathrm{COPY}_{{\cal
A}_{1}{\cal A}_{2}}U_{{\cal Q}{\cal A}_{1}}|{\psi}{\rangle}_{\cal
Q}{\otimes}|{\chi}{\rangle}_{{\cal A}_{1}}
{\otimes}|{\chi}{\rangle}_{{\cal A}_{2}} \nonumber
\\
&=&\sum_{k=1}^{{\cal D}_{\cal Q}}|x_{k}{\rangle}_{\cal
Q}{\otimes}(B_{k}|{\psi}{\rangle})_{{\cal
A}_{1}}{\otimes}|x_{k}{\rangle}_{{\cal A}_{2}}.
\end{eqnarray}
Following this unitary interaction, we carry out a projective
measurement on ${\cal A}_{1}{\cal A}_{2}$, with the projection
operators
\begin{equation}
P_{k}=1_{{\cal
A}_{1}}{\otimes}(|x_{k}{\rangle}{\langle}x_{k}|)_{{\cal A}_{2}}.
\end{equation}
The probability $P(k|{\psi})$ of the $k$th outcome is easily shown
to be ${\langle}{\psi}|{\Pi}_{k}|{\psi}{\rangle}$.  The final
state of ${\cal Q}$ is obtained by tracing the entire final state
over the ancilla.  If we write
\begin{equation}
V=\mathrm{SWAP}_{{\cal Q}{\cal A}_{1}}\mathrm{COPY}_{{\cal
A}_{1}{\cal A}_{2}}U_{{\cal Q}{\cal A}_{1}},
\end{equation}
where $V$ is clearly unitary, then when outcome $k$ is obtained
for the measurement based on the projectors $P_{k}$ in Eq. (2.39),
the state of ${\cal Q}$ is transformed by the following completely
positive, linear, trace non-increasing map:
\begin{eqnarray}
{\Phi}_{k}({\rho}_{{\cal Q}})&=&\mathrm{Tr}_{{\cal A}_{1}{\cal
A}_{2}}\left(P_{k}V({\rho}_{{\cal
Q}}{\otimes}|{\chi}{\rangle}{\langle}{\chi}|_{{\cal
A}_{1}}{\otimes}|{\chi}{\rangle}{\langle}{\chi}|_{{\cal
A}_{2}})V^{\dagger}\right) ,\nonumber \\
&=&|x_{k}{\rangle}{\langle}x_{k}|.
\end{eqnarray}
So, there is a one-to-one correspondence between the measurement
outcomes and the orthonormal states $|x_{k}{\rangle}$. This
implies that the result of the measurement is perfectly
retrodictable for an arbitrary initial quantum state.

 It is often
helpful to make use of the fact that every such map has an
operator-sum decomposition. In this case, we have
\begin{equation}
{\Phi}_{k}({\rho}_{{\cal Q}})=\sum_{r=1}^{D_{\cal
Q}}A_{kr}{\rho}_{{\cal Q}}A^{\dagger}_{kr},
\end{equation}
for some operators $A_{kr}$.  After some algebra, we find that we
may write
\begin{equation}
A_{kr}=|x_{k}{\rangle}{\langle}x_{r}|B_{k}.
\end{equation}
These are given by Eq. (2.34) if we take
\begin{equation}
B_{k}=\sum_{r=1}^{{\cal D}_{\cal
Q}}\sqrt{{\pi}_{kr}}|x_{r}{\rangle}{\langle}{\pi}_{kr}|.
\end{equation}
One can show, using Eqs. (2.32) and (2.33), that these operators
satisfy Eq. (2.35). We have thus shown how to form from a standard
implementation of a POVM  one whose outcome is perfectly
retrodictable for an arbitrary initial state.

\section{unambiguous outcome retrodiction}
\renewcommand{\theequation}{3.\arabic{equation}}
\setcounter{equation}{0}
\subsection{With entanglement}
In the preceding section, we addressed the issue of perfectly
retrodicting the outcome of a generalised measurement ${\cal
M}_{\cal Q}$ on a quantum system ${\cal Q}$ by examining the final
state when the initial state is arbitrary. Here we impose the less
stringent condition that for some known, initial state, the
outcome can always be retrodicted, unambiguously, which is to say
with zero probability of error, with some some non-zero
probability instead. We allow for the possibility that the
retrodiction attempt gives an inconclusive result.

The issues that we discuss in this subsection are insensitive to
the dimension $\tilde{D}_{\cal Q}$ of $\tilde{\cal H}_{\cal Q}$,
provided that $\tilde{D}_{\cal Q}{\geq}D_{\cal Q}$.  For maximum
generality, we should assume, and take advantage of the fact that
${\cal Q}$ can be initially entangled with some ancillary system
${\cal A}$, with corresponding Hilbert space ${\cal H}_{\cal A}$,
having finite dimension $D_{\cal A}$. These systems are initially
prepared in a joint state with corresponding density operator
${\rho}_{\cal QA}$. The measurement ${\cal M}_{\cal Q}$ is carried
out on ${\cal Q}$.  For the sake of simplicity, we will consider
only fine-grained measurements. Here, the final, normalised state
corresponding to outcome $k$ is obtained by the transformation

\begin{equation}
{\rho}_{\cal QA}{\rightarrow}{\rho}_{{\cal QA}k}
=\frac{(A_{k}{\otimes}1_{\cal A}){\rho}_{\cal
QA}(A^{\dagger}_{k}{\otimes}1_{\cal A})}{P(k|{\rho}_{\cal QA})}.
\end{equation}

Our aim is to retrodict the outcome of the measurement ${\cal
M}_{\cal Q}$ by distinguishing between the states ${\rho}_{{\cal
QA}k}$. To do this, we must perform a second measurement ${\cal
M}_{\cal QA}$ on ${\cal QA}$.  This will be tailored so that its
outcome matches that of ${\cal M}_{\cal Q}$ as closely as
possible.  As we are interested in situations where the outcome is
retrodicted unambiguously, the measurement ${\cal M}_{\cal QA}$
will have $N+1$ outcomes: $N$ of these correspond to the possible
outcomes of ${\cal M}_{\cal Q}$ and a further signals the failure
of the retrodiction attempt, making this result inconclusive.  So,
we may represent this measurement by an $N+1$-element POVM
$({\Xi}_{0},{\Xi}_{1},{\ldots},{\Xi}_{N})$ for which
\begin{equation}
\sum_{k=0}^{N}{\Xi}_{k}=1_{\cal QA}.
\end{equation}
The condition for error-free unambiguous outcome retrodiction may
be written as
\begin{equation}
\mathrm{Tr}({\Xi}_{k'}{\rho}_{{\cal
QA}k})=\mathrm{Tr}({\Xi}_{k}{\rho}_{{\cal QA}k}){\delta}_{kk'},
\end{equation}
for $k,k'{\in}\{1,{\ldots},N\}$.  The probability that the
retrodiction attempt gives an inconclusive result is
\begin{equation}
P(?|{\rho}_{\cal
QA})=\mathrm{Tr}\left({\Xi}_{0}\sum_{k=0}^{N}(A_{k}{\otimes}1_{\cal
A}){\rho}_{\cal QA}(A^{\dagger}_{k}{\otimes}1_{\cal A})\right).
\end{equation}
Under what conditions does there exist an initial state
${\rho}_{\cal QA}$ for which the outcome of the fine-grained
measurement ${\cal M}_{\cal Q}$ is unambiguously retrodictable? To
address this question, we may, without loss of generality take the
initial state to be a pure state ${\rho}_{\cal QA}=|{\psi}_{\cal
QA}{\rangle}{\langle}{\psi}_{\cal QA}|$, since any mixed state can
be purified by considering a sufficiently large ancilla ${\cal
A}$.  The Schmidt decomposition theorem implies that we can always
take the dimensionality of ${\cal H}_{\cal A}$ to be at most
$D_{\cal Q}$. We will now prove

\begin{theorem}
A sufficient condition for the existence of an initial state
$|{\psi}_{\cal QA}{\rangle}{\in}{\cal H}_{\cal QA}$ for which the
outcome of a fine-grained measurement ${\cal M}_{\cal Q}$ is
unambiguously retrodictable is that the corresponding Kraus
operators are linearly independent.  When this is the case, the
outcome of ${\cal M}_{\cal Q}$ is unambiguously retrodictable for
any known $|{\psi}_{\cal QA}{\rangle}{\in}{\cal H}_{\cal QA}$ with
maximum Schmidt rank.  When the Kraus operators are non-singular,
linear independence is also a necessary condition for the
existence of an initial state $|{\psi}_{\cal
QA}{\rangle}{\in}{\cal H}_{\cal QA}$ for which the outcome of
${\cal M}_{\cal Q}$ can be unambiguously retrodicted.
\end{theorem}
\noindent $\mathbf{Proof}$.  We will first prove necessity for
non-singular Kraus operators.  Consider the final states
\begin{equation}
|{\psi}_{{\cal QA}k}{\rangle}=P(k|{\psi}_{\cal
QA})^{-1/2}(A_{k}{\otimes}1_{\cal A})|{\psi}_{{\cal QA}}{\rangle}.
\end{equation}
If the $A_{k}$ are non-singular, then the corresponding
probabilities $P(k|{\psi}_{\cal QA})$ will be non-zero for all
$|{\psi}_{\cal QA}{\rangle}{\in}{\cal H}_{\cal QA}$.  If the
$A_{k}$ are linearly dependent, then there exist coefficients
${\alpha}_{k}$, not all of which are zero, such that
$\sum_{k}{\alpha}_{k}A_{k}=0$.  It is then a simple matter to show
that $\sum_{k}{\beta}_{k}|{\psi}_{{\cal QA}k}{\rangle}=0$, where
${\beta}_{k}={\alpha}_{k}P(k|{\psi}_{\cal QA})^{-1/2}$ and that
these are not all zero.  Hence the final states are linearly
dependent and cannot be unambiguously distinguished\cite{Melin},
so for no initial state can the outcome of the measurement ${\cal
M}_{\cal Q}$ be unambiguously retrodicted.

We now prove, again by contradiction, that linear independence of
the $A_{k}$ is a sufficient condition for being able to
unambiguously retrodict the outcome of ${\cal M}_{\cal Q}$ when
the initial state $|{\psi}_{\cal QA}{\rangle}{\in}{\cal H}_{\cal
QA}$ has maximum Schmidt rank. To do this, we make use of the fact
that linear independence of the final states is a sufficient
condition for them being amenable to unambiguous
discrimination\cite{Melin}. We write $|{\psi}_{\cal QA}{\rangle}$
in Schmidt decomposition form:
\begin{equation}
|{\psi}_{\cal QA}{\rangle}=\sum_{j=1}^{D_{\cal
Q}}c_{j}|x_{j}{\rangle}_{\cal Q}{\otimes}|y_{j}{\rangle}_{\cal A},
\end{equation}
where $\{|x_{j}{\rangle}\}$ is an orthonormal basis for ${\cal
H}_{\cal Q}$ and $\{|y_{j}{\rangle}\}$ is an orthonormal subset of
${\cal H}_{\cal A}$.  When outcome $k$ is obtained, the
post-measurement state is
\begin{equation}
|{\psi}_{{\cal QA}k}{\rangle}=P(k|{\psi}_{\cal
QA})^{-1/2}\sum_{j=1}^{D_{\cal Q}}c_{j}(A_{k}|x_{j}{\rangle}_{\cal
Q}){\otimes}|y_{j}{\rangle}_{\cal A},
\end{equation}
where the probability of outcome $k$ is
\begin{equation}
P(k|{\psi}_{\cal QA})=\sum_{j=1}^{D_{\cal
Q}}|c_{j}|^{2}{\langle}x_{j}|{\Pi}_{k}|x_{j}{\rangle}.
\end{equation}
We will assume that $|{\psi}_{\cal QA}{\rangle}$ has maximum
Schmidt  rank, that is, that all of the $c_{j}$ are non-zero. For
any initial state with this property, all of the outcome
probabilities $P(k|{\psi}_{\cal QA})$ are non-zero, even if some
of the $A_{k}$ are singular. To prove this, let $c>0$ be the
smallest of the $|c_{j}|$. Then $P(k|{\psi}_{\cal
QA}){\geq}c^{2}\sum_{j=1}^{D_{\cal
Q}}{\langle}x_{j}|{\Pi}_{k}|x_{j}{\rangle}=c^{2}\mathrm{Tr}({\Pi}_{k})$.
The ${\Pi}_{k}$ are positive operators, which, while not
necessarily being positive definite, are nevertheless non-zero.
Hence, $\mathrm{Tr}({\Pi}_{k})>0$
${\forall}\;k{\in}\{1,{\ldots},N\}$. From this, it follows that
$P(k|{\psi}_{\cal QA})>0$ ${\forall}\;k{\in}\{1,{\ldots},N\}$.

Suppose now that unambiguous outcome retrodiction is impossible,
that is, that the final states $|{\psi}_{{\cal QA}k}{\rangle}$ are
linearly dependent.  There would then exist coefficients
${\alpha}_{k}$, not all of which are zero, such that
\begin{equation}
\sum_{k=1}^{N}{\alpha}_{k}|{\psi}_{{\cal QA}k}{\rangle}=0.
\end{equation}
If we again let ${\beta}_{k}={\alpha}_{k}P(k|{\psi}_{\cal
QA})^{-1/2}$, then we see that these are not all zero and that,
with the help of Eq. (3.7), this linear dependence condition can
be written as
\begin{equation}
\sum_{k=1}^{N}{\beta}_{k}\sum_{j'=1}^{D_{\cal Q}}
c_{j'}(A_{k}|x_{j'}{\rangle}_{\cal
Q}){\otimes}|y_{j'}{\rangle}_{\cal A}=0.
\end{equation}
Taking the partial inner product of this with ${\langle}y_{j}|$
and dividing the result by $c_{j}$, we find
\begin{equation}
\sum_{k=1}^{N}{\beta}_{k}A_{k}|x_{j}{\rangle}=0\;{\forall}\;j.
\end{equation}
Finally, we make use of the completeness of the $|x_{j}{\rangle}$
and see that this, when combined with Eq. (3.11), gives
\begin{equation}
\sum_{k=1}^{N}{\beta}_{k}A_{k}=\sum_{k=1}^{N}{\beta}_{k}A_{k}\sum_{j=1}^{D_{\cal
Q}}|x_{j}{\rangle}{\langle}x_{j}|=0,
\end{equation}
that is, the $A_{k}$ must be linearly dependent.  So, for an
initial state which is pure with maximum Schmidt rank, if the
final states are unamenable to unambiguous discrimination, which
is to say that they are linearly dependent, then the Kraus
operators are also linearly dependent.  This completes the proof
.${\Box}$\\

This theorem has some interesting consequences that we shall now
describe.   The first is in relation to general quantum
operations.  These are described by completely positive, linear,
trace non-increasing maps
${\rho}{\rightarrow}{\Phi}({\rho})=\sum_{k=1}^{N}A_{k}{\rho}A^{\dagger}_{k}$,
where $\sum_{k=1}^{N}A^{\dagger}_{k}A_{k}{\leq}1_{\cal Q}$.  In a
well-known theorem, Choi\cite{Choi} showed that every such map has
an operator-sum decomposition in terms of  linearly independent
Kraus operators $A_{k}$. Combining this fact with Theorem 4, we
see that for each trace-preserving quantum operation ${\Phi}$,
there exists a fine-grained generalised measurement whose Kraus
operators form an operator-sum decomposition of ${\Phi}$ and whose
outcome is unambiguously retrodictable for all pure initial states
with maximum Schmidt rank.

A second consequence of this theorem relates to the problem of
distinguishing between unitary operators.  Childs {\em et
al}\cite{Childs} and Ac\'{\i}n\cite{Acin} have addressed the
problem of distinguishing between a pair of unitary operators.
Theorem 4 enables us to say something about the more general
problem of distinguishing between $N$ unitary operators.

The problem is this: a quantum system ${\cal Q}$ and an ancilla
${\cal A}$ are initially prepared in the possibly entangled state
${\rho}_{\cal QA}$.  With probability $p_{k}$, ${\cal Q}$ is
subjected to one of the $N$ unitary operators $U_{k}$.  The entire
state undergoes the transformation
\begin{equation}
{\rho}_{\cal QA}{\rightarrow}{\rho}_{{\cal
QA}k}=(U_{k}{\otimes}1_{\cal A}){\rho}_{\cal QA}
(U^{\dagger}_{k}{\otimes}1_{\cal A}),
\end{equation}
with probability $p_{k}$.  The aim is to determine which unitary
operator has been applied.  This is done by distinguishing between
the final states ${\rho}_{{\cal QA}k}$.

Comparison of Eq. (3.13) with Eq. (3.1) shows that this procedure
can be regarded as a particular example of retrodiction of the
outcome of a fine-grained generalised measurement, specifically
one which has the Kraus operators
\begin{equation}
A_{k}=\sqrt{p_{k}}U_{k}.
\end{equation}
Clearly, when all of the $p_{k}$ are non-zero, then linear
independence of the $A_{k}$ is equivalent to that of the $U_{k}$.
It follows from this and the non-singularity of unitary operators
that a necessary and sufficient condition for being able to
unambiguously discriminate between $N$ unitary operators $U_{k}$
for some, possibly entangled, initial state is that they are
linearly independent.

%
%-------<Moved here: Sasaki>-------
%
%      Are not there some illustrative examples for
%      a generalised measurement with non-singular Kraus operators ?
%
%  New sentense
Theorem 4 gives a special status to generalised measurements with
non-singular Kraus operators.  Measurements of this kind might
appear to be somewhat artificial constructions. After all, neither
projective measurements nor many of the optimal generalised
measurements for the various kinds of state discrimination have
this property\cite{Holevo73_POM,Helstrom_QDET,Review,Melin}.
However, it has recently been suggested by Fuchs and
Jacobs\cite{Fuchs} that such measurements may, in practice, be the
rule rather than the exception.  They argue that a measurement for
which a particular outcome is impossible to achieve for some
initial state is an idealisation that would require infinite
resources to implement (infinite precision in tuning interactions,
timings etc.) Accordingly, realistic, {\em finite-strength}
measurements do not possess this property and have non-singular
POVM elements or equivalently, for fine-grained measurements,
Kraus operators.

Of course, this reasoning also applies to the measurement which
retrodicts the outcome of ${\cal M}_{\cal Q}$.  Unambiguous
outcome retrodiction will, in general, require that the Kraus
operators of the retrodicting measurement are highly singular.
While, for the reasons given above, this is difficult, even
impossible to achieve in practice, there are, as far as we are
aware, no fundamental limitations on how well these idealised
measurements can be approximately implemented.  Finite-strength
measurements will have a special status with regard to unambiguous
outcome retrodiction if the measurement whose outcome we are
trying to retrodict is not as strong as the retrodicting
measurement.

It should also be noted that when some of the Kraus operators are
singular, linear independence is not, in general, a necessary
condition for unambiguous outcome retrodiction for some initial
state. As a counter-example, consider the case of ${\cal H}_{\cal
Q}$ being three dimensional and spanned by the orthonormal vectors
$|x{\rangle}, |y{\rangle}$ and $|z{\rangle}$. Consider now a
fine-grained measurement with the singular, linearly dependent
Kraus operators
\begin{eqnarray}
A_{1}&=&\frac{|x{\rangle}{\langle}x|}{\sqrt{2}}, \\
A_{2}&=&\frac{|y{\rangle}{\langle}y|}{\sqrt{2}}, \\
A_{3}&=&|z{\rangle}{\langle}z|, \\
A_{4}&=&\frac{|x{\rangle}{\langle}x|+|y{\rangle}{\langle}y|}{\sqrt{2}}.
\end{eqnarray}
If the initial state is $|z{\rangle}$, then we know {\em a priori}
that the only possible outcome is `3', so knowing that this state
was prepared enables us to perfectly retrodict the outcome without
having access to the measurement record.
%------------------------------
%

\subsection{Without entanglement}

The final issue we shall investigate is unambiguous outcome
retrodiction without entanglement.  For the sake of simplicity, we
again confine our attention to fine-grained measurements. When
${\cal Q}$ is initially prepared in the pure state
$|{\psi}{\rangle}$, then the final state corresponding to the
$k$th outcome is, up to a phase
\begin{equation}
|{\psi}_{k}{\rangle}=\frac{A_{k}|{\psi}{\rangle}}{\sqrt{P(k|{\psi})}},
\end{equation}
when the probability $P(k|{\psi})$ of the $k$th outcome is
non-zero. Unambiguous retrodiction of the outcome of ${\cal
M}_{\cal Q}$ with the initial state $|{\psi}{\rangle}$ is possible
only if the final states which have non-zero probability are
linearly independent. Actually, in what follows it will, for
reasons that will become apparent, be more convenient to enquire
as to when unambiguous retrodiction is {\em impossible} for every
initial state in ${\cal H}_{\cal Q}$.  Let $\sigma({\psi})$ be the
subset of $\{1,{\ldots},N\}$ for which
$A_{k}|{\psi}{\rangle}{\neq}0$ when $k{\in}\sigma({\psi})$.  Then
unambiguous retrodiction of the outcome of ${\cal M}_{\cal Q}$ is
impossible for every pure initial state
$|{\psi}{\rangle}{\in}{\cal H}_{\cal Q}$ iff there exist
coefficients ${\alpha}_{k}({\psi})$, not all of which are zero for
$k{\in}\sigma({\psi})$, such that
\begin{equation}
\left(\sum_{k{\in}\sigma({\psi})}{\alpha}_{k}({\psi})A_{k}\right)|{\psi}{\rangle}=0,
\end{equation}
for all $|{\psi}{\rangle}{\in}{\cal H}_{\cal Q}$, which is to say
iff the possible final states are linearly dependent for all
initial states.  In particular, for a finite-strength measurement,
the $A_{k}$ are non-singular and so
$\sigma({\psi})=\{1,{\ldots},N\}$.  In this case, the
impossibility condition is that for each
$|{\psi}{\rangle}{\in}{\cal H}_{\cal Q}$, there exist coefficients
${\alpha}_{k}({\psi})$, not all of which are zero, such that
\begin{equation}
\left(\sum_{k=1}^{N}{\alpha}_{k}({\psi})A_{k}\right)|{\psi}{\rangle}=0.
\end{equation}
Operators $A_{k}$ with this property are said to be {\em locally
linearly dependent}.

Locally linearly dependent sets of operators have been
investigated in detail by \v{S}emrl and
coworkers\cite{Semrl1,Semrl2}.  Notice that local linear
dependence is weaker than linear dependence, which is the special
case of the ${\alpha}_{k}$ being independent of
$|{\psi}{\rangle}$.

Equivalently, it is necessary, though not sufficient for a set of
operators to be linearly independent to not be locally linearly
dependent. Consequently, it is sufficient for the outcome of a
finite-strength, fine-grained measurement to be unambiguously
retrodictable for a single unentangled state for it to be
unambiguously retrodictable for all maximum Schmidt rank entangled
states, but not vice versa. Consider, for example,  the four,
non-singular, Pauli operators
$(1,{\sigma}_{x},{\sigma}_{y},{\sigma}_{z})$. Though linearly
independent, these operators are locally linearly dependent.   So,
as far as pure states are concerned, an entangled initial state is
required to unambiguously determine which operator has been
implemented, as in dense coding\cite{Coding}. More generally, the
non-singularity of unitary operators implies that a necessary and
sufficient condition for a set of unitary operators to be
unambiguously distinguishable with a pure, non-entangled initial
state is that they are not locally linearly dependent.

Having made the distinction between linear dependence and local
linear dependence, which is responsible for the fact that there
exist finite-strength measurements whose outcomes are
unambiguously retrodictable for some entangled but no unentangled,
pure, initial states, one particular question forces itself upon
us: given that the outcome of a measurement is unambiguously
retrodictable with an entangled, pure initial state, what
subsidiary conditions must the measurement satisfy for its outcome
to be unambiguously retrodictable for some non-entangled, pure
initial state? For finite-strength measurements, this question is
equivalent to: under what conditions is a linearly independent set
of Kraus operators, subject, of course, to the resolution of the
identity, not a locally linearly dependent set?

The problem of determining when a linearly independent set of
operators is not a locally linearly dependent set has been solved
for the special cases $N=2,3$\cite{Semrl1}. The solutions for
$N{\geq}4$ are not known at this time.  Progress has, however,
been made with regard to this problem.  It has been shown by
Bre\v{s}ar and \v{S}emrl\cite{Semrl1} that the solution for
arbitrary $N$ can be deduced from that of the problem of
classifying the maximal vector spaces of $N{\times}N$ matrices
with zero determinant.  However, this is also currently unknown.

We will examine here the solution for $N=2$ and unravel its
implications. Here, we are considering a fine-grained measurement
with two outcomes having corresponding Kraus operators $A_{1}$ and
$A_{2}$. Bre\v{s}ar and \v{S}emrl\cite{Semrl1} have shown that the
following two statements are equivalent:\\

\noindent(i) $A_{1}$ and $A_{2}$ are locally linearly
dependent.\\

\noindent(ii) (a) $A_{1}$ and $A_{2}$ are linearly
dependent or (b) there exists a vector $|{\phi}{\rangle}{\in}\tilde{\cal H}_{\cal Q}$ such that $\mathrm{span}\{A_{1}|{\psi}{\rangle}:|{\psi}{\rangle}{\in}{\cal H}_{\cal Q}\}=\mathrm{span}\{A_{2}|{\psi}{\rangle}:|{\psi}{\rangle}{\in}{\cal H}_{\cal Q}\}=\tilde{\cal H}_{\phi}$, where $\tilde{\cal H}_{\phi}$ is the one-dimensional subspace of $\tilde{\cal H}_{\cal Q}$ spanned by $|{\phi}{\rangle}$.\\

It follows that if $A_{1}$ and $A_{2}$ are linearly independent
and also locally linearly dependent, then condition (iib) must be
satisfied.  This condition, when combined  with the resolution of
the identity, implies that $D_{\cal Q}=2$ and that ${\cal H}_{\cal
Q}$ has an orthonormal basis $\{|x{\rangle},|y{\rangle}\}$ such
that
\begin{eqnarray}
A_{1}&=&|{\phi}{\rangle}{\langle}x|, \\
A_{2}&=&|{\phi}{\rangle}{\langle}y|.
\end{eqnarray}

These operators are clearly singular.  It follows that for every
two-outcome, fine-grained, finite strength measurement, if the
Kraus operators are not linearly dependent, then they are not
locally linearly dependent either.  So, for such measurements, if
the outcome can be unambiguously retrodicted for some entangled
initial state, then it can also be unambiguously retrodicted for
some non-entangled initial pure state.

 Let us conclude with an
examination of the possibility of unambiguous outcome retrodiction
for all initial states $|{\psi}{\rangle}{\in}{\cal H}_{\cal Q}$.
For a finite-strength, fine-grained measurement, the necessary and
sufficient condition for this to be possible is that for every
pure, initial state, the set of $N$ pure, post-measurement states
is a linearly independent set.  Formally, this requirement can be
expressed as
\begin{equation}
\left(\sum_{k=1}^{N}{\alpha}_{k}A_{k}\right)|{\psi}{\rangle}{\neq}0,
\end{equation}
for all non-zero $|{\psi}{\rangle}{\in}{\cal H}_{\cal Q}$ and all
complex coefficients ${\alpha}_{k}$ unless
${\alpha}_{k}=0\;{\forall}\;k\;{\in}\{1,{\ldots},N\}$.  A set of
operators $A_{k}$ with this property can be said to be
\textit{locally linearly independent}.  Local linear dependence
and local linear independence are not, like linear dependence and
independence, complementary concepts.  For example, no set of two
Pauli operators is either locally linearly dependent or locally
linearly independent.

Local linear independence is a considerably stronger condition
than linear independence.  So strong, in fact, that if ${\cal
H}_{\cal Q}$ and $\tilde{\cal H}_{\cal Q}$ are finite dimensional
and $\tilde{D}_{\cal Q}{\leq}{D}_{\cal Q}$, then it cannot be
satisfied (except in the trivial case of the equality and a
single, non-singular operator.)  It is easy to see that locally
linearly independent operators must be non-singular, so that this
condition cannot be satisfied if $\tilde{D}_{\cal Q}<{D}_{\cal
Q}$.  To prove that it cannot be satisfied when $\tilde{D}_{\cal
Q}={D}_{\cal Q}$ either, we make use of the fact that any subset
of a locally linearly independent set must also be locally
linearly independent.  Let us then consider just two operators,
$A_{1}$ and $A_{2}$. These operators must be non-singular.  This
implies, in the finite dimensional case, that if $\tilde{D}_{\cal
Q}={D}_{\cal Q}$, they must have unique left and right inverses.

Given that $A_{1}$ and $A_{2}$ are non-singular, it follows that
$A^{-1}_{1}A_{2}$ must also be non-singular.  It then has $D_{\cal
Q}$ linearly independent eigenvectors with non-zero eigenvalues.
Let ${\lambda}{\neq}0$ be an eigenvalue of $A^{-1}_{1}A_{2}$ with
corresponding eigenvector $|{\psi}{\rangle}$.  Now consider
\begin{equation}
A^{-1}_{1}(-{\lambda}A_{1}+A_{2})|{\psi}{\rangle}=(-{\lambda}+{\lambda})|{\psi}{\rangle}=0.
\end{equation}
Operating throughout this equation with $A_{1}$, we find that
\begin{equation}
(-{\lambda}A_{1}+A_{2})|{\psi}{\rangle}=0,
\end{equation}
and so the operators $A_{1}$ and $A_{2}$ cannot be locally
linearly independent.  From this, it follows that, for finite
dimensional quantum systems, if the dimension of the output
Hilbert space does not exceed that of the input Hilbert space,
then fine-grained measurements with locally linearly independent
Kraus operators are impossible. However, one can devise examples
of such measurements for finite dimensional quantum systems if the
output Hilbert space has higher dimension than the input Hilbert
space. Let ${D}_{\cal Q}=2$ and $\tilde{D}_{\cal Q}=4$. Also, let
$\{|x_{1}{\rangle},|x_{2}{\rangle}\}$ and
$\{|\tilde{x}_{1}{\rangle},|\tilde{x}_{2}{\rangle},|\tilde{x}_{3}{\rangle},|\tilde{x}_{4}{\rangle}\}$
be orthonormal basis sets for ${\cal H}_{\cal Q}$ and $\tilde{\cal
H }_{\cal Q}$ respectively.  Consider now a two-outcome,
fine-grained measurement whose Kraus operators have the following
matrix representations in these bases:
\begin{equation}
A_{1}=\frac{1}{\sqrt{2}} \left(
\begin{array}{cc} 1&  0 \\ 0&
1 \\ 0&  0 \\ 0& 0
\end{array}
\right),\;\;\;\;A_{2}=\frac{1}{\sqrt{2}} \left(
\begin{array}{cc} 0&  0 \\ 0& 0 \\1&  0 \\ 0&
1
\end{array}
\right),
\end{equation}
that is, the row-$j'$, column-$j$ element of $A_{k}$ is
${\langle}\tilde{x}_{j'}|A_{k}|x_{j}{\rangle}$.  One can easily
verify that $A^{\dagger}_{1}A_{1}+A^{\dagger}_{2}A_{2}=1_{\cal
Q}$, and so these operators constitute a fine-grained measurement.
To prove that they are locally linearly independent, let us write
an arbitrary pure initial state in ${\cal H}_{\cal Q}$ as
$|{\psi}{\rangle}=c_{1}|x_{1}{\rangle}+c_{2}|x_{2}{\rangle}$.
Then,
\begin{eqnarray}
&&({\alpha}_{1}A_{1}+{\alpha}_{2}A_{2})|{\psi}{\rangle}
\nonumber \\
&=&\frac{c_{1}({\alpha}_{1}|\tilde{x}_{1}{\rangle}+{\alpha}_{2}|\tilde{x}_{2}{\rangle})+c_{2}({\alpha}_{1}|\tilde{x}_{3}{\rangle}+{\alpha}_{2}|\tilde{x}_{4}{\rangle})}{\sqrt{2}}.
\end{eqnarray}
As a consequence of the orthonormality of the
$|\tilde{x}_{j'}{\rangle}$, when either or both $c_{1}$ and
$c_{2}$ are non-zero, this expression is never equal to the zero
vector unless ${\alpha}_{1}$ and ${\alpha}_{2}$ are equal to 0.
Hence, the operators $A_{1}$ and $A_{2}$ are locally linearly
independent. In fact, these operators satisfy the condition in Eq.
(2.9) for {\em perfect} retrodiction for an arbitrary initial
state condition in ${\cal H}_{\cal Q}$.

There also exist interesting examples of measurements with locally
linearly independent Kraus operators on infinite dimensional
quantum systems.  Consider a bosonic mode with Hilbert space
spanned by the orthonormal occupation number states $|n{\rangle}$,
$n=0,1,2,{\ldots}$.  Now consider a two-outcome generalised
measurement with the Kraus operators
$A_{1}={\mu}\sum_{n=0}^{\infty}|n+1{\rangle}{\langle}n|$ and
$A_{2}=\sqrt{1-|{\mu}|^{2}}\sum_{n=0}^{\infty}|n{\rangle}{\langle}
n|$, where $0<|{\mu}|<1$.  It is a simple matter to show that
$A^{\dagger}_{1}A_{1}+A^{\dagger}_{2}A_{2}=\sum_{n=0}^{\infty}|n{\rangle}{\langle}n|=1$,
so that these operators do indeed form a fine-grained generalised
measurement. To show that these operators are locally linearly
independent, let the initial state of the system be the pure state
$|{\psi}{\rangle}=\sum_{n=0}^{\infty}c_{n}|n{\rangle}$, where at
least one of the $c_{n}$ is non-zero. The operators $A_{1}$ and
$A_{2}$ will be locally linearly independent iff, for every such
state, and for every pair of complex coefficients ${\alpha}_{1}$
and ${\alpha}_{2}$, at least one of which is non-zero,
\begin{equation}
({\alpha}_{1}A_{1}+{\alpha}_{2}A_{2})|{\psi}{\rangle}{\neq}0.
\end{equation}
To show that this is so, let $n_{0}$ be the smallest value of $n$
for which $c_{n}{\neq}0$.  It follows then that
${\langle}n_{0}|A_{1}|{\psi}{\rangle}=0$ and
${\langle}n_{0}|A_{2}|{\psi}{\rangle}=\sqrt{1-|{\mu}|^{2}}c_{n_{0}}$.
Hence,
\begin{equation}
{\langle}n_{0}|({\alpha}_{1}A_{1}+{\alpha}_{2}A_{2})|{\psi}{\rangle}={\alpha}_{2}\sqrt{1-|{\mu}|^{2}}c_{n_{0}}
\end{equation}
which is non-zero for non-zero ${\alpha}_{2}$, implying that when
${\alpha}_{2}{\neq}0$, (3.29) is satisfied.  To show that it is
also satisfied when ${\alpha}_{2}=0$, we simply make use of the
fact that if this were not the case, then we would have
$A_{1}|{\psi}{\rangle}=0$, which is not true. We can see this, for
example, by making use of the fact that
${\langle}n_{0}+1|A_{1}|{\psi}{\rangle}={\mu}c_{n_{0}}{\neq}0$.

The key property which makes the operators $A_{1}$ and $A_{2}$
defined above a locally linearly independent set is the fact that
$A_{1}$ has no eigenvalues/eigenvectors. In fact, it is
straightforward to prove that, for any pair of non-singular
operators $A_{1}$ and $A_{2}$, if $A_{2}$ is proportional to the
identity, then local linear independence
 of $A_{1}$ and $A_{2}$ is equivalent to the condition that $A_{1}$ has no eigenvalues/eigenvectors\cite{footnote2}.

\section*{Discussion}

In this paper, we have addressed the following problem:  suppose
that a generalised measurement has been carried out on a quantum
system.  We do not know the outcome of the measurement.  We do,
however, know which measurement has been carried out and have
access to the system following the measurement.  We are free to
interrogate the final state in any way which is physically
possible. Our aim is to devise a suitable `retrodicting'
measurement which will reveal the outcome of the first
measurement.

This task is simple if the initial measurement is a projective
measurement; if there  no irreversible evolution following this
measurement, then we can simply reverse any evolution that occurs
and perform the same measurement again. Generalised measurements
do not, however, possess the repeatability property which is
responsible for the straightforward nature of outcome retrodiction
for projective measurements.  In section II, we derived a
necessary and sufficient condition on the Kraus transformations
operators for the outcome of a generalised measurement to be
perfectly retrodictable for an arbitrary initial state.  We also
showed that there is no advantage to be gained if the initial
state, though arbitrary, is known.

When the input and output Hilbert spaces have the same dimension,
the only fine-grained measurements  which satisfy this condition
are projective measurements, possibly followed by an
outcome-independent unitary transformation. We also showed that
every POVM can be realised by a measurement whose outcome is
perfectly retrodictable for all initial states iff the number of
outcomes does not exceed the output Hilbert space dimension.  We
also described an algorithm by which such an implementation can be
constructed using a standard implementation.  This essentially
involves swapping the information contained in the measuring
apparatus and the system following the measurement.

We then addressed the problem of unambiguously retrodicting the
outcome of a generalised measurement, with zero probability of
error but with a possible non-zero probability of the retrodiction
attempt giving an inconclusive result.    We addressed this issue
in section III, focusing on fine-grained measurements.  The fact
that only linearly independent pure, final states can be
unambiguously discriminated places constraints on the Kraus
operators of such measurements.  We showed that if entanglement
with an ancillary system is possible, then a sufficient and, for
finite-strength measurements, necessary condition is that the
Kraus operators are linearly independent. This result has
interesting connections with a theorem due to Choi and also with
the problem of unambiguously discriminating between unitary
operators.

When the initial state is pure and entanglement is not permitted,
we have shown that the issue of unambiguous outcome retrodiction
is closely related to the concepts of operator local linear
dependence and local linear independence. While being interesting
in their own right, our demonstration that these concepts are
relevant to quantum measurement theory gives a further incentive
to explore them.

\section*{Acknowledgements}

We would like to thank Peter \v{S}emrl and Leo Livshits for
helpful correspondence and Stephen M. Barnett for illuminating
discussions about retrodiction.  We also thank Osamu Hirota, Jerry
Finkelstein and Paul Busch for suggesting helpful clarifications.
This work was supported by the UK Engineering and Physical
Sciences Research Council and by the British Council.


\begin{thebibliography}{99}
\bibitem{footnote1}
This is not, in general, true for measurements which have an
infinite number of potential outcomes.  Consider, for example, the
POVM with elements
${\Pi}({\phi})=\mathrm{const.}{\times}|{\phi}{\rangle}{\langle}{\phi}|$,
where $|{\phi}{\rangle}$ varies over all normalised states
$|{\phi}{\rangle}{\in}{\cal H}_{\cal Q}$ and the resolution of the
identity is formed by integration with respect to the appropriate
Haar measure and with an appropriate choice of the multiplicative
constant. Clearly, for any particular state $|{\chi}{\rangle}$, we
can find a state $|{\phi}{\rangle}$ which is orthogonal to it. It
follows that we have
$P({\phi}|{\chi})={\langle}{\chi}|{\Pi}({\phi})|{\chi}{\rangle}=0$.
\bibitem{footnote2}
If $({\alpha}_{1}A_{1}+{\alpha}_{2}A_{2})|{\psi}{\rangle}=0$, then
${\alpha}_{1}{\neq}0$ and
$A_{1}|{\psi}{\rangle}=-({\alpha}_{2}/{\alpha}_{1})\sqrt{1-|{\mu}|^{2}}|{\psi}{\rangle}$
and so $|{\psi}{\rangle}$ is an eigenvector of $A_{1}$ with
eigenvalue $-({\alpha}_{2}/{\alpha}_{1})\sqrt{1-|{\mu}|^{2}}$.
Conversely, if $A_{1}|{\psi}{\rangle}={\lambda}|{\psi}{\rangle}$
then
$[A_{1}-({\lambda}/\sqrt{1-|{\mu}|^{2}})A_{2}]|{\psi}{\rangle}=0$
and so $A_{1}$ and $A_{2}$ are not locally linearly independent.
\end{thebibliography}
\end{document}